\documentclass[10pt]{article}
\usepackage{graphicx} 

\usepackage{epsfig}
\usepackage{latexsym}
 \usepackage{amsmath}

\begin{document}

\title{\Large \bf Hurst  exponent of very long birth time series in XX century Romania. \\Social and religious aspects }

\author{ \large \bf G. Rotundo$^{1,}$\footnote {giulia.rotundo@gmail.com,} ,  M. Ausloos$^{2,3,4}$\footnote {marcel.ausloos@ulg.ac.be,ma683@le.ac.uk} ,    C. Herteliu$^{5,}$\footnote {claudiu.herteliu@gmail.com} , B. Ileanu$^{5, }$\footnote {ileanub@yahoo.com}  
 \\ \\$^1$  Sapienza University of Rome, Faculty of Economics,
\\Department of Methods and models for Economics, Territory and Finance,
\\via del Castro Laurenziano 9, I-00161 Roma, Italia 
\\ $^2$ GRAPES, rue de la Belle Jardiniere, B-4031 Liege, \\Federation Wallonie-Bruxelles, Belgium
\\ $^3$ e-Humanities Group, KNAW,\\Joan Muyskenweg 25, 1096 CJ Amsterdam, The Netherlands \\ $^4$ School of Management, University of Leicester, \\
 University Road, Leicester, LE1 7RH,  UK  
\\$^5$ The Bucharest University of Economic Studies,  Bucharest, Romania \\ }

\maketitle

  \begin{abstract}
 The Hurst exponent of very long birth time series in Romania has been extracted from official daily records, i.e. over 97 years between 1905 and 2001 included. The series  result from distinguishing between families located in urban (U) or rural  (R) areas, and belonging (Ox) or not  (NOx) to the orthodox  religion.  Four  time series  combining both   criteria, (U,R) and (Ox, NOx), are   also examined.  
 
 A statistical information is given on  these sub-populations measuring their XX-th century state as a snapshot.   However, the main goal is  to investigate whether the "daily"  production of babies is purely noisy or is fluctuating according to some non trivial fractional Brownian motion, - in  the four types of  populations, characterized by either  their habitat or  their religious attitude, yet living within the same political regime.  One of the goals was also to find whether  combined criteria implied a different behavior.  Moreover, we wish to observe whether some seasonal   periodicity exists.

The detrended fluctuation analysis technique is used for finding the fractal correlation dimension of  such  (9) signals. It has been  first necessary, due to two periodic tendencies, to define the range regime in which the Hurst exponent is  meaningfully defined.   It results that the birth of babies in all cases is a very strongly persistent signal. It is found that   the signal fractal  correlation dimension is  weaker (i) for  NOx than  for  Ox, and (ii) or U with respect to R. Moreover, it is observed that the combination of  U or  R with NOx   or OX enhances the UNOx, UOx,  and ROx fluctuations, but  smoothens the   RNOx signal, thereby  suggesting a  stronger conditioning on religiosity rituals or rules.  
\end{abstract}
 \maketitle

\section{  	Introduction }
 
 The  detrended fluctuation analysis (DFA) method  \cite{DNADFA} is  commonly used to sort out the characteristic Hurst exponent of  time series, or its fractal dimension.  Alas, many  series are    of quite finite size. However, the  daily birth evolution in Romania during the XX century is known,  from official surveys. It amounts to  about 35 600 data points, - which an interesting  very long time series.  
 
  More precisely,  the  available time series  goes from   Jan.  01, 1905  till  Dec. 21, 2001, i.e. for 97 years, thus  35 429  days; see Sect. \ref{data}. Moreover,  the birth records distinguish between several  social characteristics.   In particular, we consider  here below the  aspects of a urban (U) $vs.$ rural (R)  family location, and the orthodox (Ox)   $vs.$ nonorthodox  (NOx) family ground. Note that it might be of interest to distinguish cases within the NOx population, i.e. between various religions or sects.   Indeed, it has been shown that religious values and practices have some influenced on pregnancy outcomes \cite{SSM26.88.401musp_najman}.   In the present case, there are not enough data points for a valuable statistics when considering these sub-populations. Whence we regrouped all non orthodox affiliations into NOx.  However, we also study the combination of such "degrees of freedom" location and religiosity. Information on such population distributions in Romania is given in  an Appendix (Appendix  A) for completeness.   We emphasize that the goal is not to report a snapshot of the baby production in Romania in the last century but to examine whether some modelisation  is realistic, e.g., through a  non-trivial fractional Brownian motion.
  
In other words, a goal is  to investigate whether the "daily"  production of babies is purely noisy or is fluctuating, -  in four types of  populations, characterized by either  their habitat or  their religious attitude, yet living within the same political regime.
  For the present paper we focus on considerations, on whether there is a coherent or not behavior in baby births, when either the family location or  some family religiosity implies a different collective behavior, and whether  the combined set of  these two "parameters" (of family characteristics) leads to similar or different outputs.  The answer is searched for on whether the various time series correspond to a  fractional Brownian Motion   \cite{west,addison,falconer}, characterized by  its fractal dimension  or its so called Hurst exponent  \cite{Hu5}.

   Another goal was to find whether some periodicity exists, e.g. a seasonal one rather than a yearly one. The  DFA is useful in this respect, see Sect. \ref{comment}.

 On the methodology side, recall that the  DFA  technique \cite{DNADFA} is often used to study the correlations in the
fluctuations of stochastic time series.   It  has been much used in econophysics in the recent past     \cite{4,3a,3b}, but in many other fields as well, as e.g.  in meteorology  \cite{kima2}, in real signal spectral analysis  \cite{JAM.02.ICSALA} or faked ones \cite{PhA384.07FPMAGR}.
Briefly speaking,   the DFA technique consists in dividing a signal time series $y(t_i)$ of
 $N$ points, measured at discrete regularly spaced (or not)  instants $t_i$  into an integer number of boxes $\simeq N/\tau$ having an equal size ($=n$), such that  these non-overlapping boxes (called also windows), each
contain  $\tau$ points \cite{DNADFA}. The local trend $z(t_i)$ in each box is
 assumed  to be linear, for simplicity.  From the ordinate of a linear least-square fit of the data points in
that box, the detrended fluctuation function, $\phi^2(\tau)$, is then calculated following:

\begin{equation} \label{DFAeqFn}
\phi^2(\tau) = {1 \over \tau } {\sum_{i=k\tau+1}^{(k+1)\tau}
{\left[y(t_i)- z(t_i)\right]}^2} \qquad k=0,1,2,\dots,\left(\frac{N}{\tau}-1\right).
\end{equation}

Averaging $\phi^2(\tau)$ over the $N/\tau$ intervals gives the mean-square
fluctuations

\begin{equation}  \label{DFAeqalpha}
F(n)=< \phi^2(n)>^{1/2} \sim n^{\alpha} 
 \end{equation}

The exponent $\alpha$,  is  considered to be identical to the Hurst exponent 
when the data
is stationary; see Sect. \ref{comment} for a comment on the presently investigated case.   For the reader's own reflexion, it can be here recalled that  the  (Hurst) exponent of a time dependent signal  not only represents the so called    {\it roughness} of the signal but also the behavior of the auto-correlation function $c (\tau)$ 
 \begin{equation}\label{corr}
c (\tau) = {\langle {| y(t_{i+r})-y(t_i) |} \rangle}_{\tau}\;,\end{equation}
therefore   expected to behave like $ \sim{\tau}^{Hu}$  
\cite{Hu5}.

 The output  ($\alpha$)  implies the existence of long-range
correlations when $\alpha$  is not a half-odd integer.   In fact, $\alpha$ is an accurate measure of the most characteristic (maximum) dimension of a  fractal process \cite{kiandma2}.
 An
exponent $\alpha>1/2$ implies a  {\it  persistent} (smooth) behavior, while $\alpha<1/2$ means
a so-called {\it anti-persistent} (rough) signal \cite{west}. The latter  regime implies a signal fractal dimension $D$ close to 2, the  former close to 1. Obviously, the signal fractal   (correlation) dimension: $D_2 = 2- \alpha$, where $\alpha$  is the slope of the  $F(n)$  trend, on a  $F(n)$ $vs$. $n$ log-log plot.. The simple Brownian motion is thus characterized by $\alpha=1/2$ and white noise by $\alpha=0$ \cite{west}. Practically, the  characteristics fractional Brownian motion values are found to
lie between $0$ and $1$ \cite{4,3a,3b}.   Our findings are given in Sect. \ref{discussion}.

\section{Data}\label{data}

 Data was provided by the Romanian National Institute of Statistics (NIS) from the 2002 and 1992 censuses. We have used a query tool available within NIS intranet web site ($http://happy:81/PHC$) regarding the 2002 and 1992 censuses. (A simplified (on a 10\% sample) version for that tool is available for anyone  at $http://colectaredate.insse.ro/phc/public.do?siteLang=en$.) 
The analyzed data represents the number of babies born in Romania and which are still alive at the census in 1992 and in 2002. Data  from  Jan. 01, 1905 till Dec. 31, 1991 (31776 points) is obtained from the  1992 census; the   data from 1992 to 2001 from  the 2002 census contains 3653 points.

Two 5 years time interval examples, extracted from such time series, are shown in Fig. \ref{plot163765copy}, on a reasonable scale for better visualization. They correspond to the birth number of  babies in $urban$ areas between either Jan. 01, 1937 and Jan. 01,  1942, on one hand (red dots) and between Jan. 01, 1965 and Jan. 01, 1970, on the other hand (green dots).

These examples have been  selected for  showing  a state of "nothing  too special", although at the WWII time and one "very special" time interval, illustrated by  a much relevant event in Romania history, corresponding to    Ceausescu decree ($\#$770) forbidding abortion, on Oct. 01, 1966. A remarkable peak is observed in the data in the following year. Note that after that decree  
the cohort born in 1967 doubled compare to the previous year by  almost 1 000  babies.

\section{Technical points}\label{comment}

\begin{figure}
\centering
 \includegraphics [height=12.5cm,width=12.5cm]{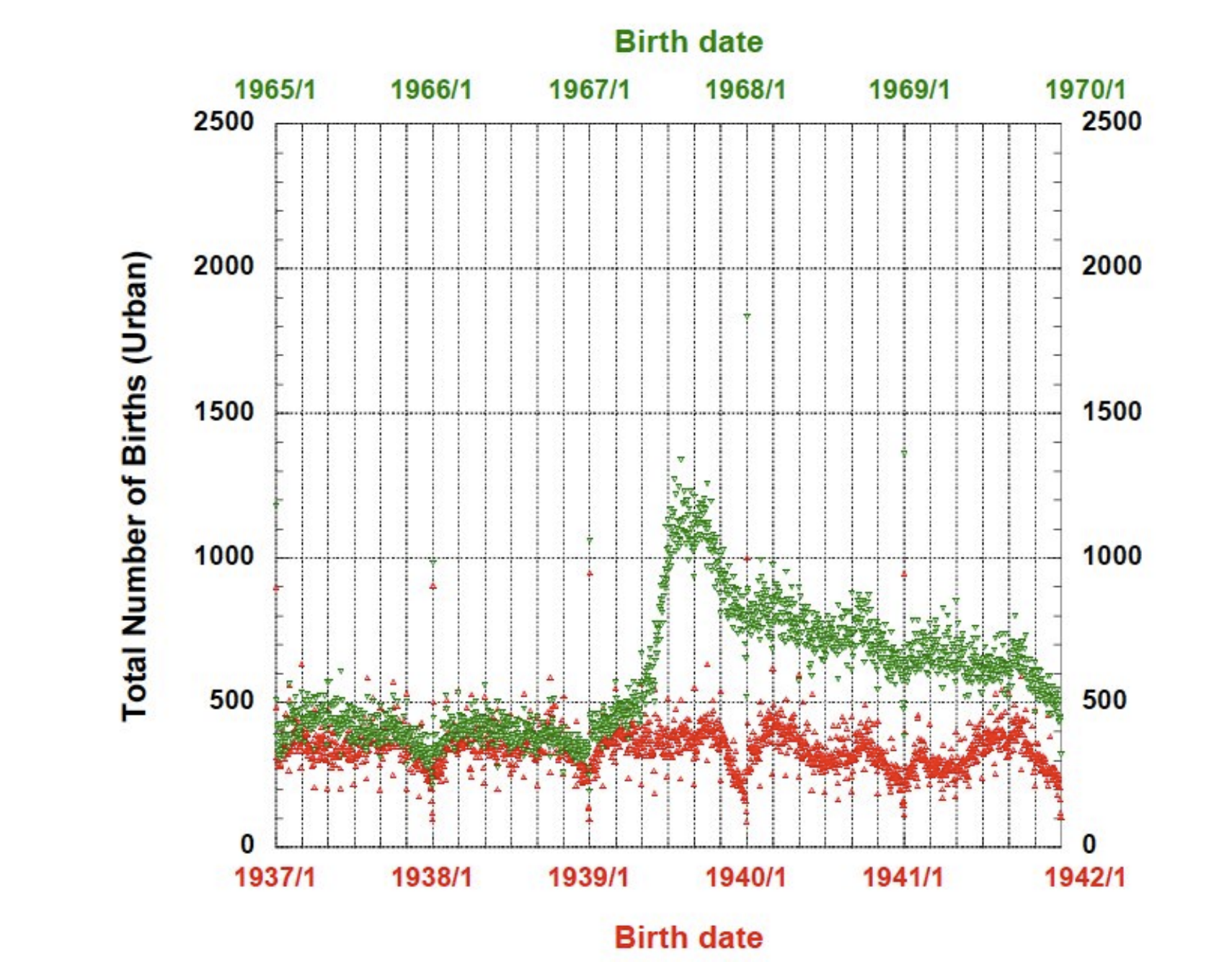}
 \caption   {  
Number of babies born  from families in urban locations illustrated as a  time series in WWII time [1937-1942] and near Ceausescu forbidding  illegal abortion decree  (Oct.1, 1966) [1965-1970]}   \label{plot163765copy} 
\end{figure}

\begin{figure}
\centering
 \includegraphics [height=10.5cm,width=12.5cm]{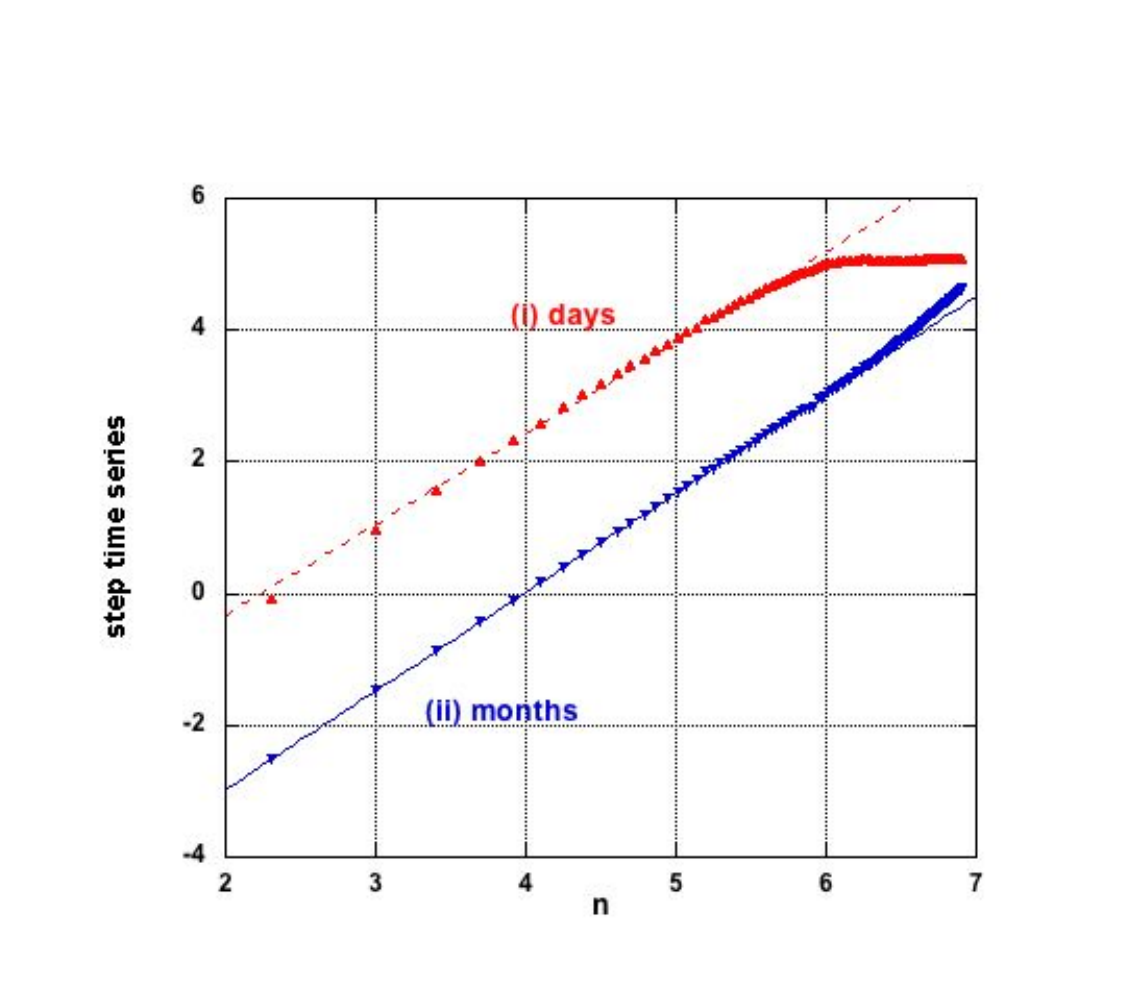}
  \caption   {  
DFA analysis of  a staircase function series mimicking the "year" as a function of time, (i) days or (ii) months; a change in slope (from 1.4  to $\sim$ 0) can be  observed at  $n=$ 5.9 for (i), corresponding to $\sim$ 365 days  }\label{ScreenPlot23daymonthsteps} 
\end{figure}

\begin{figure}
\centering
 \includegraphics [height=12.5cm,width=12.5cm]{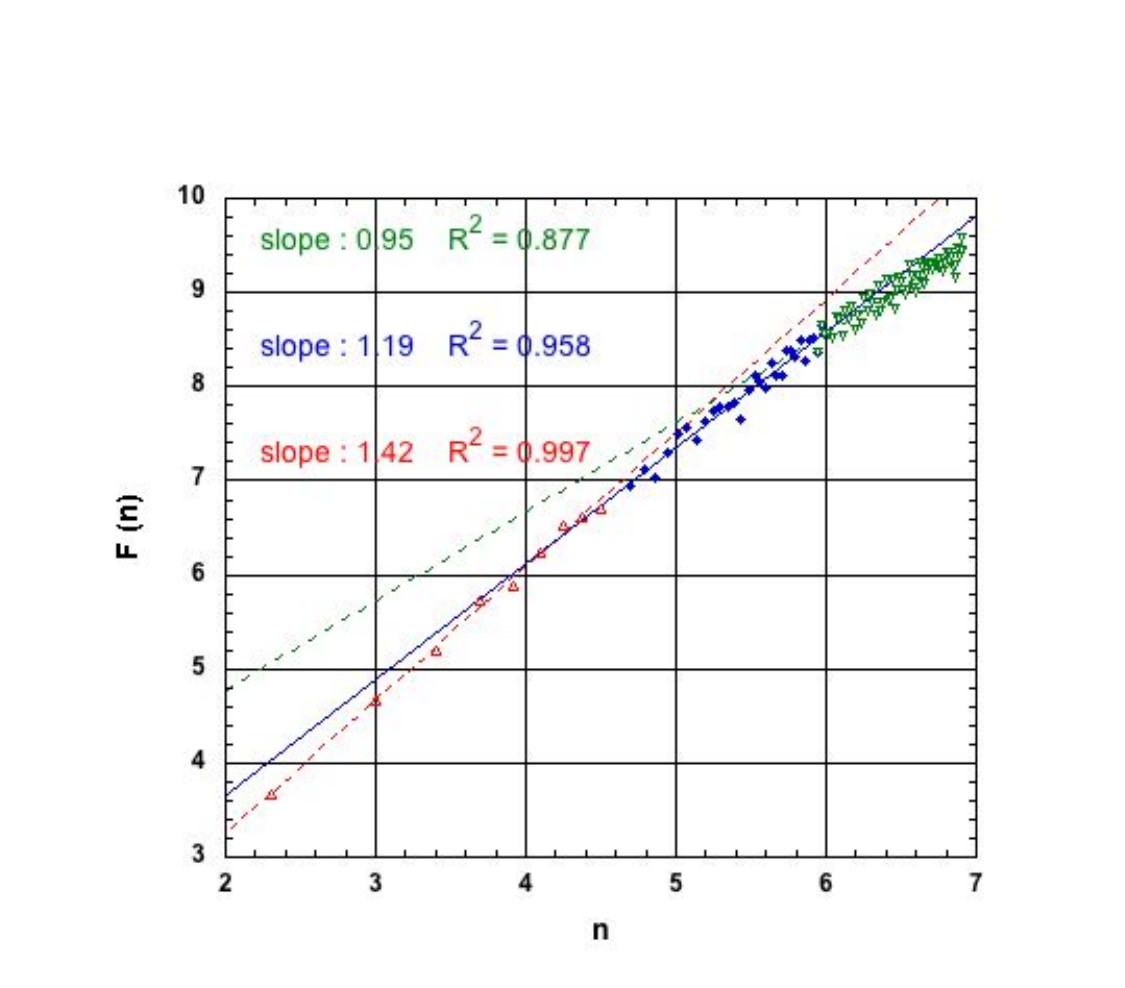}
  \caption   {  
Log-log plot illustrating the DFA  result about the  yearly periodicity of the  daily Temperature in a Romania location (see text) over the 35 429 data points; the 3 fits pertain to different  $n$ (box size)  regimes: high ($n\ge 6$), medium $n$ and low ($n\le4.5$) ranges respectively }\label{DFAT3fits} 
\end{figure}

Several  technical points are in order. 
\begin{itemize}
\item
  
Regression lines,  to obtain $\alpha$ in Eq.(\ref{DFAeqalpha}), were obtained  through  the maximum likelihood method \cite{MLE01,Seal52}  in order to avoid biases as observed in the  least-square regressions in the log-log domain \cite{SIAM51.09.661powerlaws}.  For a discussion of  finite sample effects in sequence analysis,  e.g. see   \cite{ebeling280}, but it can be reminded from a technical point of view that Wheaton et al. \cite{Wheaton95} showed that the least squares 2 free parameter  fitting  processes can be equivalent to the maximum likelihood  method when Poisson statistics apply.
 
For completeness, note that  other methods for fitting to the power-law distributions, as expected here, 
 provide biased estimates for the power-law exponent \cite{0402322fittingtopowerlaw_v3}.  

\item 
 Note that 35 429 is almost a prime number; it can only be decomposed into the product 71 x 499; which would mean to have only two boxes for calculating $F(n)$.  Therefore, during the DFA procedure, i.e. before decomposing the data series into  equal size boxes (before searching for the   residuals and averaging, as described in Eq.(\ref{DFAeqFn}), several data points are sometimes "not considered" in the process, in order to maintain an equal number of points in each box. 
 
 Taking all this into account, the $n$ range which has been always examined goes from $\sim 6.91$ ($\sim$ 1000 days) down to $\sim
2.30$,  ($\sim$ 10 days), -  resulting in a series of 100  different box sizes $n$.
 
\item

Note that one should be aware of  some possible yearly or seasonal periodicity in  the data:  it is of common knowledge that  babies are not equally born on each day of a year (see Appendix B).  Thus some regularity might be expected, though with amplitude fluctuations over the considered time interval, about one century.  On the other hand, it is of common knowledge  that every calendar year in Romania starts (in the Gregorian calendar)  on Jan. 01, and has "usually" 365 days. There are  24 occurrences of a leap  year, but the implications seem  marginal.

Every year, there are also 12 months in a year.  This is  roughly seen  on Fig. \ref{plot163765copy}.   It has been shown by Hu et al. \cite{PRE64.01.011114DFAhu_0103018}  that a periodicity in a time series implies a break (a change in slope) in the  $F(n)$ plot at the $n$ value  corresponding to the  period. Thus,  a "break in slope" at 365 days $\rightarrow$n= 5.90 $\simeq ln(365)$ has to be expected in the subsequent plots.  Moreover, seasonality  might be a sub-period constraint, i.e. 3 months (or $\simeq$ 90 days) $\rightarrow$ $n$=4.5 ($\simeq$ $ln$(90))  or  even 4 months (or $\simeq$ 120 days) $\rightarrow$ $n$=4.8 ($\simeq$ $ln$ (122)).  To  demonstrate  the possible periodicity in the present investigation,
we have imagined a time series with a constantly increasing   amplitude, i.e.  adding  (i) a constant  step each day over 365 or 366 days, depending on the year, like a series of step functions leading to a regular staircase, or (ii)  a constant step during  each month (with various days)  of the 97 years  of interest,   leading also to a staircase.  Fig. \ref{ScreenPlot23daymonthsteps}   shows  the periodicity effect in the (i)  and (ii) time series on a $F(n)$ $vs.$ $n$ log-log plot.   A well marked slope break, from 1.4  to $\sim$ 0, is obvious at $n\simeq 5.9$ for the (i) time series,$\rightarrow \sim $  365 days.
A break in slope is also well observed at  $n=$ 5.9   for the (ii) time series. This, together with Hu et al.  \cite{PRE64.01.011114DFAhu_0103018} considerations, suggests an upper range analysis  limit  to be $n\le6.0$. The lower range limit is less obvious.

Another $independent$ illustration test of  a seasonal periodic trend can be performed on  some average  daily temperature, in order to observe the lower $n$ range limit.  For these,   our source is  the Romanian National Meteorological Agency\footnote{  via $http://www.ecad.eu/dailydata/customquery.php$}. We chose to test the DFA application to the  temperature data series, pertinent to Bucharest Airport. This is illustrated in Fig. \ref{DFAT3fits}. Three power law regimes are outlined. Two, the extreme regimes,  large $n$ and low $n$, are giving absurd results. Thereby the lower range limit of interest  is  found to be at $n=4.5$ ($\rightarrow \sim 90$ days,  corresponding to a trimester. 

Thus,  only the DFA data points between  $n=$ 4.5  and  5.9 should be taken into account for measuring  a meaningful Hurst exponent. Therefore, this  periodicity elimination criterion has been used for  defining a valid  time window range in the birth data series analysis. 

\item Moreover, a  question is  often raised for statistical purposes, i.e. whether the data is stationary or not \cite{Maddalabook}. This theoretical question seems somewhat practically irrelevant in  many cases, like finance, meteorology,  and demography, because the data is "obviously" never stationary. A restricted criterion on  stationarity  seems likely sufficient for discussion and scientific progress:  if the data mean $and$ the
whatever-extracted-parameter do not change too much,  allowing for a moderate trend, as a function of time,   the data can be  called quasi-stationary.  Nevertheless the matter can be considered thereafter to be the source of a fundamental investigation. The prefix ''quasi'' is in fact, practically, conventionally dropped. Not having {\it a posteriori} noticed any specific effect which could be called spurious or anomalous, we have considered  our study and findings worth of report, according to the standards  mentioned here above.
 
Therefore, we have made some stationarity test  \cite{Dickey,Perron,KPSS}. All series seem to be stationary in this respect. 
Results of these tests can be provided if needed, but are not displayed in order  not to overload the present report.
\end{itemize}

 \begin{figure}
\centering
 \includegraphics [height=10.5cm,width=12.5cm] {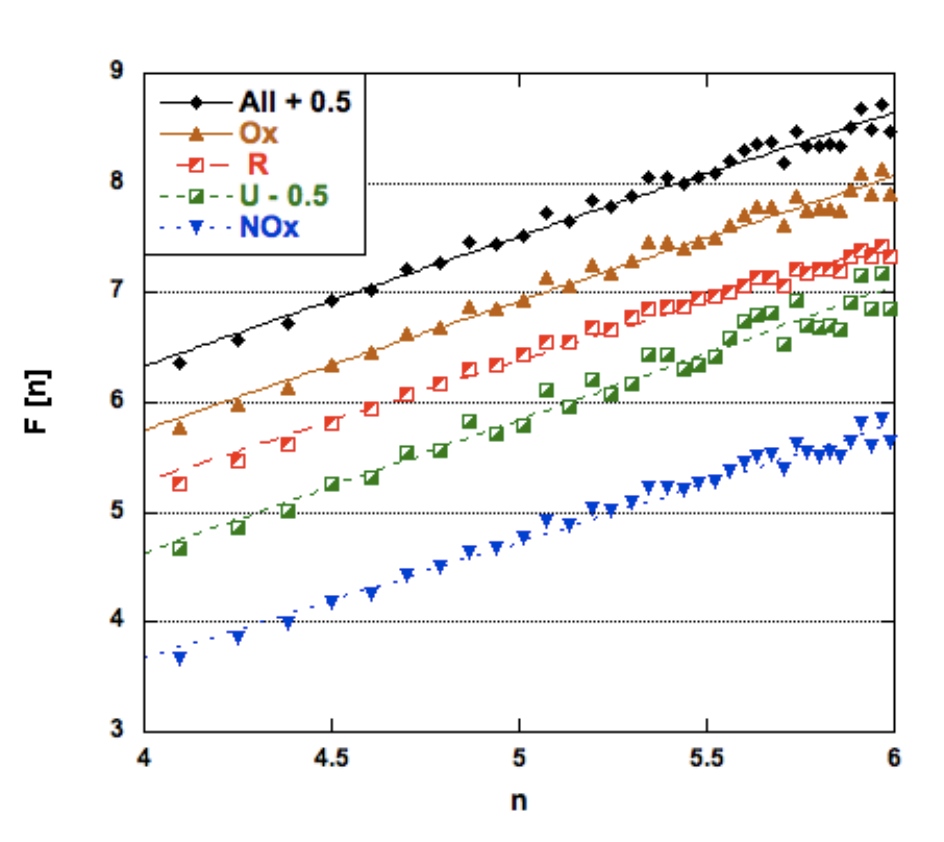}
 \caption   {  
   DFA for  the daily number of babies born  from either Orthodox  (Ox)   (triangle tip up) or Non-orthodox (NOx) (triangle tip down) Romanian  families   and of the number of babies born form Romanian families living in either rural (R) (upper left filled square) or urban (U)  (lower  right filled square)  locations in the XX-th century. For readability, the U   F(n)   amplitude has been displaced; so is the all family case (All) (diamond). 
  }   \label{Plot5} 
\end{figure}

   \begin{figure}
\centering
 \includegraphics[height=12.5cm,width=12.5cm] {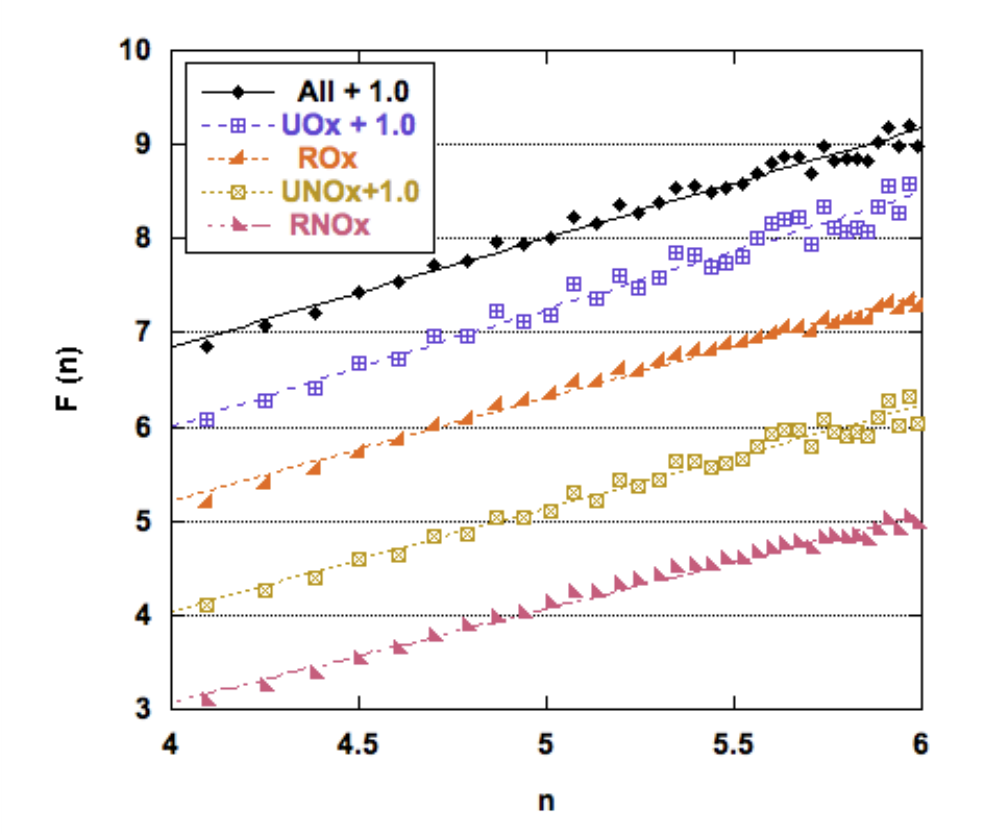}
 \caption   {      DFA for  the daily number of babies born  from Orthodox  (Ox) or Non-orthodox (NOx) Romanian  families  living in either rural (R) or urban (U) locations in the XX-th century, i.e. ROx (tip of rectangular triangle to the right), RNOx (tip of rectangular triangle to the left),  UOx (+  in square), and UNOx  (X-cross in square). For readability, both  UOx and UNOx  F(n)   amplitudes have been displaced; so is the all family case (All) (diamond). }
   \label{Plot6} 
\end{figure}
\section{Discussion}\label{discussion}

The enclosed figures, Figs. \ref{Plot5}-\ref{Plot6}, provide the relevant results of the present study. They report successively the best fits of the relevant $F(n)$ function on a log-log plot and the deduced slope, corresponding to $\alpha$ and equivalently to the Hurst exponent in the following cases, together with the regression coefficient:

\begin{itemize}
\item  (All) number of babies born  from  all Romanian families  over most of the XX-th century: Figs. \ref{Plot5}-\ref{Plot6}, 
\item (U) number of babies born  from  Romanian families in $urban$   locations: Fig. \ref{Plot5},
\item (R) number of babies born  from Romanian  families in $rural$ locations: Fig. \ref{Plot5},
\item (Ox) number of babies born  in Orthodox  families: Fig. \ref{Plot5},
\item (NOx) number of babies born  in  Non-orthodox (NOx) Romanian  families: Fig. \ref{Plot5};
\item (UOx) number of babies born  from Orthodox  families in $urban$ locations: Fig. \ref{Plot6}, 
\item (UNOx) number of babies born  from Non-Orthodox  families in  $urban$  locations: Fig. \ref{Plot6}, 
\item (ROx) number of babies born  from Orthodox  families in  $rural$  locations: Fig. \ref{Plot6}, 
\item (RNOX) number of babies born  from Non-Orthodox  families in  $rural$  locations: Fig. \ref{Plot6}.
\end{itemize}

All  $\alpha$ values  for  these different cases are summarized in Table \ref{Table1}. It can be observed that the regression coefficient is  quite large, often much larger than 0.96 for the one-criterion  filtered data, and above 0.92 for the 2-criteria filtered data..   The Hurst exponent  usually has  either a high value ($\sim$0.9) close to  that of a strict   {\it persistent} signal \cite{west}, or is near 1.2, - mostly for the 2-criteria filtered signals.   
For  comparing with a visual case, let it be known that the roughness of the  line of summits in the French Alps near Chamonix, FR, has a fractal dimension = 1.2.

 It can be observed that some very slight difference exists between Ox and NOx; it might be partially due to   the amplitude values, as well as between  rural or   urban  locations, for the same reason. In such cases,  the signal fractal dimension is always weaker for NOx than Ox, and for R with respect to U.
 
 The most interesting point is found when observing the fractal dimension or Hurst exponent when two criteria are imposed: the  birth in UOx  (urban \& orthodox) families  becomes less coherent ($D_2\ge 1.2$), while RNOx (rural \& non-orthodox) families) tends toward a  smoother persistent behavior.

 \begin{table} \begin{center}
 \begin{tabular}[t]{|c|c|c|c|c|c|c|c|c|c| }
   \hline
                                       &All      	&U            &R      	& Ox    	& NOx.   			&  UOx     &UNOx. 	&ROx    	&RNOx \\ \hline
  $D_2$                     &1.132 	&1.041	& 1.146 	&1.166	& 1.127  			& 1.227	&1.206  	&1.295	&1.077  \\\hline
    $\alpha$                &0.821 	&0.959	& 0.854 	&0.834	& 0.873  			& 0.773	&0.794  	&0.705	&0.923  \\
$\pm \delta\alpha$ &0.021 	&0.034	& 0.016 	&0.022	& 0.053  			& 0.046	&0.056  	&0.033	&0.057  \\
 $R^2$                       &0.980 	&0.964	& 0.989	&0.980	&  0.934			& 0.927	& 0.916	&0.957	&0.934\\ \hline
  \hline
&Figs. \ref{Plot5}-\ref{Plot6} &   \multicolumn{4}{|c|}{Fig. \ref{Plot5}}  &    \multicolumn{4}{|c|}{Fig. \ref{Plot6} } 

 \\\hline
\end{tabular} 
 \caption{ Resulting values of the   fits  to the indicated data with  a power law, Eq.(\ref{DFAeqalpha}), for obtaining the Hurst exponent  $\alpha$  and its error $\delta\alpha$,  from which the signal fractal dimension, 
 $ D_2=2-\alpha$, can be obtained. }\label{Table1}
 \end{center}
 \end{table}

 \section{ Conclusions} \label{sec:conclusions}
 
This  report has presented a study of the behaviors of populations according to long time series data.  One of the goals  was to investigate whether the "daily"  production of babies  is a purely noisy  signal or is fluctuating according to some non trivial fractional Brownian motion. We investigated   various types of  populations, characterized by either  their habitat or  their religious attitude, yet living within the same political regime. We also 
searched  whether some periodicity exists in such  signals.  

We describe the data and the methodology.  We could analyze very long time series giving the number of babies born per day in Romania during the  XX century. We used a simple DFA method on all together 9 cases.  In conclusion, it  has been  found that  beyond the signal periodicity and the various historical and political events, the correlations between births lead to a very persistent signal, - whatever the  family location or religious   membership, even though the data looks very stochastic.

We found  collective coherence. However, we have also shown  a combined criterion effect.  The Non-orthodox sub-population living in urban areas,   has a behavior similar to that of the Orthodox population, wherever it is located, but the RNOx population departs from  a  usual  random process, - for us implying a  stronger conditioning on religiosity rituals or rules.   
 
 For further studies, we may  suggest (as  also recommended by a reviewer) to go beyond the simple DFA, in order to investigate  what power-law cross correlations  can be found between  such  simultaneously recorded time series, thus generalizing Eq.(\ref{corr}).   To do so, the Detrended Cross-Correlation Analysis  (DCCA) \cite{DCCAPRL10094HES,DCCAEPL94HES,DCCADCDIS18Dong,DCCAAPPB43Wang},   also including multifractality  aspects \cite{muDCCAEPJB72Jafari,muDCCAPhA389Hajian,muDCCAJTP321Stan,muDCCAPRE89Drozdz}, can be  useful.   However, this   leads to  different aspects than those hereby  searched for and outlined.

 \begin{flushleft}
{\large \bf Acknowledgment}
\end{flushleft}
 This paper is part of  GR and MA scientific activities in COST Action TD1210   'Analyzing the dynamics of information and knowledge landscapes'.

\vskip0.5cm

 \begin{flushleft}
{\large \bf Appendix A: Statistical summary on population communities}
\end{flushleft}

A few  general information details  about population ethnicity, structure by urban/rural, gender, religion, in Romania could serve to enlighten considerations by the reader.

The  most important community,  from  the religion point of view,  according to the census sources  is the Eastern-Orthodox (86.8\%). The  so called   "non-orthodox families"  (13.2\%) are mainly made of Roman-Catholics (4.7\%), Reformations (3.2\%), Pentecostals (1.5\%), Greek-Catholics (0.9\%), Baptists (0.6\%), Seventh Day Adventists (0.4\%),  Moslems (0.3\%), Unitarians (0.3 \%), Lutherans (0.3\%), Evangelicals (0.2\%) and  Old rite Christians (0.2\%).  Other denominations, including atheists,  have  each smaller a size.

From the geographical point of view the ratio between rural and urban population was $ \simeq 0.8708$ $ ca.$  2000.  It has evolved with time, of course. Indeed, the urbanity dramatically increased in the 5-6th decade of the last century with forced industrialization imposed by communists.  Many villages  were  suddenly declared as   towns.  

Moreover, from the ethnic point of view the most important  communities, according to the 2002 Census official data, are: Romanians (89.5\%), Hungarians(6.6\%), Rromans ($\equiv$ Gypsies) (2.5\%), Ukrainians(0.3\%), Germans (0.3\%), Russians (0.2\%), Turks (0.2\%), Tatars(0.1\%) and Serbians (0.1\%). 
 
 The intersection of such sets can be described in a statistical way through distribution characteristics; see Table \ref{Romanianstatselect}.  Only a few selected groups are mentioned. Notations seem obvious. More information can be obtained from the authors if necessary, upon request. 

\begin{table} \begin{center}
  \begin{tabular}{|c|c|c|c|c|c|c|c|c|c|c|c|c|c|}
   \hline
 &  min  & Max &Sum&mean & med.  &Std. Dev.&Skewn.&Kurt. \\\hline
	all  U	&	5	&	2158	&	13335045	&	376.388	&	358 	&	206.39 	&	0.371 	&	0.880	\\
	all U rmn	&	3	&	1910	&	11604268	&	327.536	&	306 &	185.44 	&	0.405 	&	0.737 	\\
	all U hung	&	0	&	78	&	360793	&	10.184	&	3 &	12.95&	1.308	&	0.811 	\\
	all U gyps	&	0	&	35	&	197990	&	5.588 	&	4 	&	5.49 	&	0.965 	&	0.268 	\\
	all U germ	&	0	&	11	&	32560	&	0.919 	&	0 	&	1.31	&	1.805	&	3.743	\\
	\hline
all  R	&	14	&	1383	&	11612016	&	327.755	&	336 	&	130.94 	&	0.022 	&	0.835 	\\
	all R rmn	&	9	&	1228	&	10018082	&	282.765  	&	289 &	114.28 	&	0.076	&	0.754 	\\
	all R hung	&	0	&	85	&	328882	&	9.283	&	2	&	11.20 	&	0.885 &	-0.358	\\
	all R gyps	&	0	&	49	&	296861	&	8.379 	&	5 	&	8.14 	&	1.018 	&	0.164	\\
	all R germ	&	0	&	9	&	15818	&	0.447	&	0 	&	0.83 	&	2.277	&	6.150	\\
\hline
	all Ox U 	&	3	&	1944	&	11578538	&	326.810	&	307 &	186.96 	&	0.411	&	0.854	\\
	  Ox U rmn	&	3	&	1992	&	11672383	&	329.458 	&	314 	&	189.35	&	0.433 	&	1.043 	\\
	 Ox U hung	&	0	&	145	&	625732	&	17.663	&	18 	&	18.28 	&	0.490 	&	-0.718 	\\
	 Ox U gyps	&	0	&	31	&	191381	&	5.402	&	4 	&	4.90	&	0.784 	&	-0.0087	\\
	 Ox U germ	&	0	&	16	&	55800	&	1.575	&	1 	&	1.97 &	1.284	&	1.245 \\
\hline
	 all Ox R 	&	8	&	1267	&	10051258	&	283.701 	&	289 	&	116.22	&	0.089	&	0.862	\\
	 Ox R rmn	&	8	&	1274	&	10019369	&	282.801 	&	282 	&	117.66	&	0.204	&	1.069 	\\
	 Ox R hung	&	0	&	113	&	457184	&	12.904 	&	12 	&	13.54	&	0.515 	&	-0.613	\\
	 Ox R gyps	&	0	&	42	&	280903	&	7.929	&	6 &	6.83&	0.765	&	-0.166	\\
	 Ox R germ	&	0	&	9	&	25832	&	0.729 	&	0 	&	1.16	&	1.865 	&	3.786 	\\
\hline 
\end{tabular}
\caption{A few statistical characteristics of the daily number of births in Romania   during the XX-th century according to different criteria  for characterizing   families, not only location and religiosity, but also  ethnicity }\label{Romanianstatselect}
 \end{center}
 \end{table}
 
    \begin{figure}
\centering
 \includegraphics[height=13.5cm,width=12.5cm]  {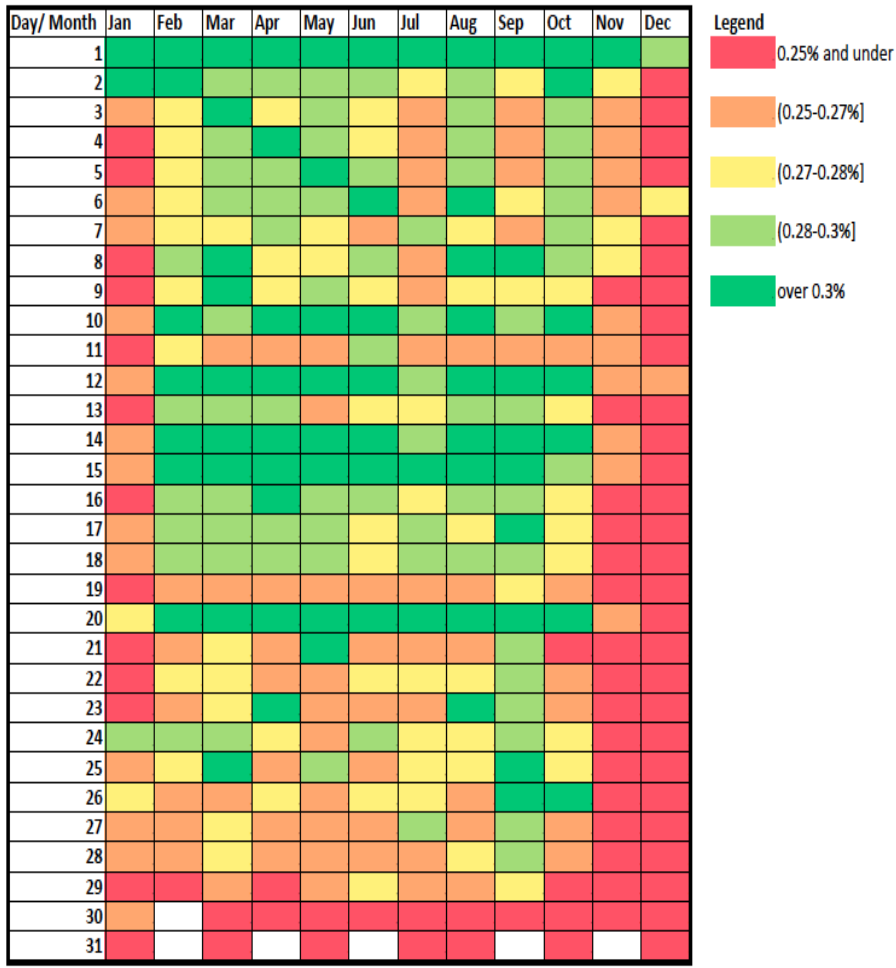} 
 \caption   {      Table of the daily percentage level of births for the  97  years studied in the main text, showing seasonal effects. }
   \label{97_years_births} 
\end{figure}

\vskip0.5cm   \begin{flushleft}
{\large \bf Appendix B:   Seasonal effects on births}\label{appseasonal}
\end{flushleft}

 Babies are not regularly born during the years.  The phenomenon has always existed, but the peaks have evolved with time, due to socio-economic, political and religiosity conditions. It is known that  the phenomenon mainly depended  on seasons, because of night duration and temperature.       Wargentin \cite{Wargentin} showed that the baby production was more frequent in december than in other months; he suggested  "causes" for his finding. Quetelet \cite{Quetelet} found a birth peak in February  and a minimum in July for Belgium  and The Netherlands, in [1815-1826].  Moheau  \cite{Moheau} already distinguished country side from cities,  but found in both cases, a birth peak in spring time, thus a  summer production.production. Villerm\' e  \cite{Villerme} pretended that the cause for such variations was the temperature.  Nowadays, i.e. in present times, the temperature cause  has to be  much disregarded \cite{ popsoc474.11.1season}.
 
 The seasonality effect is less marked nowadays in advanced civilizations. Nevertheless, it is still observed, e.g. in Romania, as shown on Fig. \ref{97_years_births}.   The figure results from a daily aggregation  for all 97 years,  whence the data concerns  almost 25  10$^6$ births.  After aggregating the number of births data for each day, the daily shares were calculated. Next    the   values  were coded within their quintiles; different colors are used to distinguish  different groups. It can be understood that  a  depletion  during wither months  (December-February) implies a lack of production at the end of winter and beginning of spring, when  work activity in the fields has to resume.   Recall  that  a uniform distribution requires for each day a 1/365.25  share, i.e. $\simeq
  0.2732785\%.$ N.B. We have applied a $\chi^2$ test  for comparing to a uniform distribution and  found statistically significant differences: $ p \le$ 0.001.

 Notice that we have  investigated, as a possible "parameter",  some  religious affiliation which implies some specific sexual behavior at various times during the year, thus leading to something else that a seasonal effect.  It is known that religious rituals are quite tied to seasonal activities. Therefore, a weaker  attachment to religious principles might  modify  the  seasonality effect \cite{ popsoc474.11.1season}. However, a Fourier transform of the examined  data in the main text and an extraction of whatever cycles are not part of the present investigations, only concerned by whether the daily birth production is a coherent or not signal.

\end{document}